\documentstyle[11pt]{article}
\begin{document}

\def    \beq    {\begin{equation}} \def \eeq    {\end{equation}}
\def    \bea    {\begin{eqnarray}} \def \eea    {\end{eqnarray}}
\newcommand\mx[4]{\left#1\begin{array}{#2}#3\end{array}\right#4}

\begin{center}\Large \textbf{A note on half-supersymmetric bound states in 
M-theory and type IIA}

\end{center}

\vspace{0.3cm}

\begin{center}

\textbf{Henric Larsson}

\normalsize
\vspace{0.2cm}
\emph{Institute for Theoretical Physics \\
G\"oteborg University and Chalmers University of Technology \\
SE-412 96 G\"oteborg, Sweden \\
E-mail: solo@fy.chalmers.se}

\end{center}
\vspace{0.5cm}
\begin{center}

\large {\bf Abstract} \end{center} \small
By using $O(7,7)$ transformations, to deform D$6$--branes, we obtain 
half-supersymmetric bound state solutions of type IIA supergravity,
containing D6, D4, D2, D0, F1-branes and waves. We lift the solutions
 to M-theory which gives half-supersymmetric M-theory bound states, e.g.  
KK6--M5--M5--M5--M2--M2--M2--MW. We also take near horizon limits for
 the type IIA 
solutions, which gives supergravity duals of 7--dimensional 
 non-commutative open string theory 
 (with space-time and space-space non--commutativity), non--commutative 
Yang-Mills theory (with space-space and light-like non--commutativity) and 
an open D4--brane theory.

\vspace{0.5cm}

\normalsize
\section{Introduction}
Many different dual supergravity solutions corresponding to bound states have
 been constructed 
\cite{sugra1,cederwall1,has}, using different solution generating methods 
\cite{sugra1,cederwall1,has,per,solo}. 
Most of these bound states have D3, D4, D5, NS5 or M5--branes as the 
highest rank brane.
 These bound states are of interest since they are useful in the 
investigation of non--commutative Yang-Mills theory (NCYM) 
\cite{has,NCYM,al,solo},   
non--commutative open string theory (NCOS) \cite{sei,NCOS,solo},
OM/ODp/$\widetilde{\rm ODq}$ theory \cite{OM,ODp,al1,ODq,solo1} and 
wound string theory \cite{WST}. More 
specifically, they are used to construct supergravity duals of 
non--commutative open brane (or gauge) theories, by taking appropriate near 
horizon limits. E.g., for a bound state which has been obtained by an NS-NS 
$B_{01}$ deformation of a D$p$--brane, one has to take an `electric' near 
horizon limit, while
 for a bound state with $B_{12}$ one has to take a `magnetic' near horizon 
limit. In the first case one obtains the supergravity dual of 
$(p+1)$-dimensional space-time non--commutative open string theory, 
which can be seen 
by investigating the properties of 
the open string quantities (see e.g. \cite{solo})
\begin{equation}\label{NCOS}
\frac{G_{\mu\nu}}{\alpha'}=\frac{1}{\alpha'_{\rm eff}}H^{-1/2}\eta_{\mu\nu}\ ,
\quad 
G^{2}_{OS}=gH^{\frac{3-p}{4}}\ ,\quad \Theta^{01}=\alpha'_{\rm eff}\ ,
\end{equation}
where $G_{\mu\nu}$ is the open string metric, $G^{2}_{OS}$ is the open string 
coupling constant and $\Theta^{01}$ is the non--commutativity parameter.
 Note that the open string metric and coupling constant approach constant 
values in the decoupling limit $\frac{r}{R}\rightarrow\infty$, since 
$H=1+\Big(\frac{R}{r}\Big)^{7-p}\rightarrow 1$. This implies that 
the tension of open strings, in the 
decoupling limit, is $T=1/\alpha'_{\rm eff}$, which means that massive open 
string modes are kept in the spectrum. 
 In the second case one obtains the supergravity dual of non--commutative 
super-Yang-Mills, with $\Theta^{12}\neq 0$ and not an open string theory, since
the massive open strings decouple.
For a more detailed introduction to supergravity duals and the definition
 of `electric' and `magnetic' near horizon limits, see Section 
3.2 below and \cite{solo,solo1}.

In this note, the main interest is to construct supergravity solutions, 
corresponding to half-supersymmetric bound states, for which 
the D6--brane (IIA) or the KK6--brane (M-theory) is the brane of highest 
dimensionality\footnote{For D6--brane bound states with lower supersymmetry, 
see e.g. \cite{sato}.}. All these bound states are 
half-supersymmetric since we start with a half-supersymmetric D6--brane and 
use $O(7,7)$ transformations to obtain the bound states (more specifically we 
only use T-duality and gauge transformations, see Section 2 and 
\cite{per,solo} for details). We also construct a half-supersymmetric type 
IIB supergravity solution, corresponding to a 
D7--D5--D5--D5--D3--D3--D3--D1--F1 bound state (see Appendix A). This bound 
state is included since it can not be obtained from the 
D6--brane bound states, by using T-duality. 

The next step, towards constructing all possible supergravity solutions, 
corresponding to M-theory and IIA/B bound states, would be to construct bound
 states where the D8--brane is the brane of highest dimensionality. However, 
since the 
D8--brane only exists in massive type IIA supergravity, one would have to 
adjust the deformation procedure that is used in this note (see Section 2),
 in order to be able to deform
 the D8--brane. Bound states containing D8--branes will not be discussed 
further in this note (except in Section 5 Discussion).  

As an application, we take near horizon limits of
 the type IIA bound states and obtain supergravity duals of NCOS (with 
space-time 
and space-space non--commutativity), NCYM (with space-space and light-like 
non--commutativity) and an open D4--brane theory. 

This note is organized as follows: In Section 2 we give an introduction to 
the most important features of 
the solution generating technique that we will use in Section 3 and 4. 
In the next Section we construct various bound 
states containing D6--branes and lower rank D--branes as well as F-strings 
and waves. We also take various near horizon limits, which give supergravity 
duals of different non--commutative open brane (or gauge) theories.  
 In Section 4 we lift the type IIA bound state solutions to 11 
dimensions, which gives half-supersymmetric (since uplifting to 11 
dimensions preserves supersymmetry) bound states containing 
KK6, M5, M2--branes and M-waves. We conclude with section 5 which is 
discussion. 

\section{Deformation of D$p$--branes}

The solution generating technique that we will use was derived in
\cite{per} (see also \cite{solo}) following earlier work in \cite{has}. For
 an NS deformation of a
 general D$p$--brane one first T-dualizes 
in the directions where one wants to turn on NS fluxes, and then one
shifts $B_2$ with a constant in these directions. After this
one T-dualizes back again. In a more precise 
language, the
deformation with parameter $\theta^{\mu\nu}$ is generated by
the following $O(p+1,p+1)$ T-duality group element\footnote{See 
\cite{per,solo} for conventions and definitions of the 
various elements of $O(p+1,p+1)$ appearing in the following discussion. Note 
that in this note $\theta^{\mu\nu}$ is dimensionless while in 
\cite{per,solo,solo1}, 
$\frac{\theta^{\mu\nu}}{\alpha'}$ were dimensionless.}

\beq \Lambda=\Lambda_0\dots
\Lambda_p\Lambda_{\theta}\Lambda_p\cdots \Lambda_0=
J\Lambda_{\theta}J=\Lambda_{-\theta}^T =
\mx{(}{ll}{1&0\\ \theta&1}{)}\ ,  \eeq

where $\theta^{\mu\nu}$ is dimensionless and carries indices
upstairs since it starts life on the T-dual world volume \cite{per}.

Starting with a D$p$--brane solution\footnote{We are using
 multi-form notation such that $C$ is a sum of forms while $B$ (see below)
 has fixed rank $2$. Note also that in this example the 
exact form of e.g. $g_{\mu\nu}$ and $e^{2\phi}$ is not important.}

\begin{eqnarray}
ds^{2}&=&g_{\mu\nu}dx^{\mu}dx^{\nu}+g_{ij}dy^{i}dy^{j}\ ,\quad e^{2\phi}\ , 
\nonumber\\
C&=&\omega dx^{0}\wedge \cdots \wedge dx^{p}+\gamma_{7-p}\ ,
\end{eqnarray}
where $x^{\mu}$, $\mu=0,\ldots ,p$, are coordinates in the brane directions 
while $x^{i}$, $i=p+1,\ldots ,9$, are coordinates in the transverse 
directions.
 $\gamma_{7-p}$ is a transverse form, i.e., $i_{\mu}\gamma_{7-p}=0$, where 
$i_{\mu}$ denotes the inner derivative in the $\mu$ direction.   
Now following the above deformation procedure gives the following deformed 
configuration \cite{solo}:
\begin{eqnarray}\label{def}
\tilde{g}_{\mu\nu}&=&g_{\mu\rho}\Big[(1-(\theta g)^{2})^{-1}\Big]^{\rho}_{}
{\nu}
\ , \quad \tilde{g}_{ij}=g_{ij}\ , \nonumber\\
\tilde{B}_{\mu\nu}&=&-g_{\mu\rho}\theta^{\rho\sigma}g_{\sigma\lambda}
\Big[(1-(\theta g)^{2})^{-1}\Big]^{\lambda}_{}{\nu}\ ,\\
e^{2\tilde{\phi}}&=&\frac{e^{2\phi}}{\sqrt{{\rm det}(1-(\theta g)^{2})}}=
e^{2\phi}\Big(\frac{{\rm det}\tilde{g}}{{\rm det}g}\Big)^{\frac{1}{2}}\ ,
\nonumber\\
\tilde{C}&=&e^{-\frac{1}{2}\tilde{B}_{\mu\nu}dx^{\mu}\wedge dx^{\nu}}\big(
\omega e^{\frac{1}{2}\theta^{\mu\nu}i_{\mu}i_{\nu}}dx^{0}\wedge\cdots \wedge 
dx^{p}+\gamma_{7-p}\big)\ .\nonumber
\end{eqnarray}

There are two types of deformations that are possible: $\theta^{0i}$ and 
$\theta^{ij}$, where $i,j=1,2,\ldots,p$. The first one is called `electric' 
since we mix the time direction with a spatial direction,
while the second is called `magnetic' since the time direction is not included.

Electric deformations are used if one would like to include F1--strings in the 
bound state, while magnetic deformations are used to include D$q$--branes, 
where $q=p-2,p-4,\ldots,p-2n$, and $n$ is the number of magnetic deformations
(i.e., one turn on $\theta^{12},\theta^{34}\ldots,\theta^{2n-1,2n}$).

To include waves, one has to mix an electric and a magnetic deformation with 
equal `strength', i.e., turn on e.g. $\theta^{01}$ and $\theta^{12}$ where 
$\theta^{01}=\pm \theta^{12}$. 

\section{Type IIA bound states and non--commutative theories}

\subsection{Type IIA bound states}
In this section, we will apply the above deformation procedure to the 
D$6$--brane. The undeformed, half-supersymmetric,
D$6$--brane configuration is given by

\begin{eqnarray}
ds^2&=&H^{-1/2}(-(dx_{0})^{2}+\ldots +(dx_{6})^{2}) + H^{1/2}(dr^{2}+r^{2}
d\Omega^{2}_{2})\ ,\nonumber\\
C&=&\frac{1}{gH}dx^0\wedge\cdots dx^6+\frac{1}{g}R\epsilon_{1}\ ,\\ 
e^{2\phi}&=&g^{2}H^{-3/2}\ ,\quad H=1+\frac{R}{r}\ , \quad 
R=gN\sqrt{\alpha'}\ ,\nonumber
\end{eqnarray}

where $d\epsilon_{1}$ is the volume form of the 2-sphere and $g$ is the 
closed string coupling constant. We will now apply
 an $O(7,7)$ transformation on a D6--brane (see above and \cite{per} for more
details) in order to obtain bound states containing D$6$--branes and lower 
dimensional D--branes and possibly F-strings and waves.  

Deforming the D6-brane by turning on $\theta^{01}$, $\theta^{12}$,
 $\theta^{34}$ and $\theta^{56}$, gives the following deformed 
configuration:\footnote{For $\theta^{01}=0$ the metric, 
NS-NS 2-form, dilaton and the RR 1- and 3-forms have been obtained 
in \cite{al,oz}, using a different solution generating technique.} 

\begin{eqnarray} \label{D6W}
d\tilde{s}^{2}&=&\frac{H^{-1/2}}{h}(-h_{12}(dx^{0})^{2}+(dx^{1})^{2}
+h_{01}(dx^{2})^{2})-2\frac{H^{-3/2}}{h}\theta^{01}\theta^{12}
dx^{0}dx^{2}\nonumber\\
& &+H^{-1/2}\Big(\frac{1}{h_{34}}((dx^{3})^{2}+(dx^{4})^{2})+\frac{1}{h_{56}}
((dx^{5})^{2}+(dx^{6})^{2})\Big)\nonumber\\
& &+H^{1/2}(dr^{2}+r^{2}d\Omega_{2}^{2})\ ,\nonumber\\
e^{2\tilde{\phi}}&=&g^{2}\frac{H^{-3/2}}
{hh_{34}h_{56}} \ ,\nonumber\\
\tilde{B}&=&{\theta^{01}\over Hh}dx^0\wedge dx^1
-{\theta^{12}\over  Hh}dx^1\wedge dx^2
-{\theta^{34}\over  Hh_{34}}dx^3\wedge dx^4
-{\theta^{56}\over  Hh_{56}}dx^5\wedge dx^6\ ,\nonumber\\
g\tilde{C}_7&=&{1\over Hhh_{34}h_{56}}dx^0\wedge\cdots \wedge 
dx^6+\frac{\theta^{34}\theta^{56}}{H^{3}hh_{34}h_{56}}\Big(\theta^{12}dx^{1}
\wedge dx^{2}\\
& &-\theta^{01}dx^{0}\wedge dx^{1}\Big)\wedge dx^{3}
\cdots dx^{6}\wedge R\epsilon_{1}
\ ,\nonumber\\
g\tilde{C}_5&=& -\frac{\theta^{56}}{Hhh_{34}}dx^{0}\wedge 
\cdots \wedge dx^{4}-\frac{\theta^{34}}
{Hhh_{56}}dx^{0}\wedge dx^{1}\wedge dx^{2}\wedge dx^{5}\wedge dx^{6}
\nonumber\\
& &-\frac{1}{Hh_{34}h_{56}}\Big(\theta^{01}
dx^{2}\wedge \cdots \wedge dx^{6}+\theta^{12}dx^{0}\wedge
 dx^{3}\wedge dx^{4}\wedge dx^{5}\wedge dx^{6}\Big)\nonumber\\
& &+\frac{1}{H^{2}}\Big(\frac{\theta^{12}\theta^{34}}
{hh_{34}} dx^{1}\wedge dx^{2}\wedge dx^{3}\wedge dx^{4}
+\frac{\theta^{12}\theta^{56}}{hh_{56}} dx^{1}
\wedge dx^{2}\wedge dx^{5}\wedge dx^{6}\nonumber\\
& &+\frac{\theta^{34}\theta^{56}}{h_{34}h_{56}} dx^{3}
\wedge dx^{4}\wedge dx^{5}\wedge dx^{6}-\frac{\theta^{34}\theta^{01}}
{h_{34}h} dx^{0}
\wedge dx^{1}\wedge dx^{3}\wedge dx^{4}\nonumber\\
& &-\frac{\theta^{56}\theta^{01}}{h_{56}h} dx^{0}
\wedge dx^{1}\wedge dx^{5}\wedge dx^{6}\Big)\wedge R\epsilon_{1}\ ,\nonumber\\
g\tilde{C}_{3}&=&\frac{\theta^{34}\theta^{56}}{Hh} dx^{0}
\wedge dx^{1}\wedge dx^{2}+\frac{\theta^{12}\theta^{56}}{Hh_{34}} dx^{0}
\wedge dx^{3}\wedge dx^{4}+\frac{\theta^{12}\theta^{34}}{Hh_{56}} dx^{0}
\wedge dx^{5}\wedge dx^{6}\nonumber\\
& &+\frac{\theta^{01}\theta^{34}}{Hh_{56}} dx^{2}\wedge dx^{5}\wedge 
dx^{6}+\frac{\theta^{01}\theta^{56}}
{Hh_{34}} dx^{2}\wedge dx^{3}\wedge dx^{4}-\tilde{B}\wedge R\epsilon_{1}
\ ,\nonumber\\
g\tilde{C}_{1}&=&R\epsilon_{1}-\frac{\theta^{12}\theta^{34}\theta^{56}}{H}
dx^{0}-\frac{\theta^{01}\theta^{34}\theta^{56}}{H}dx^{2}\ ,\nonumber \\
\mbox{with}& & h_{i,i+1}=1+(\theta^{i,i+1})^{2}H^{-1}\
 , \quad i=1,3,5\ ,\quad h_{01}=1-(\theta^{01})^{2}H^{-1}\ ,\nonumber\\
& & h=1-(\theta^{01})^{2}H^{-1}+(\theta^{12})^{2}H^{-1}\ .\nonumber
\end{eqnarray}

This solution gives rise to many different bound states depending on the 
values 
of the $\theta$ (deformation) parameters. If 
$\theta^{01}=\pm \theta^{12}\neq 0$ there is a wave included in the 
bound state. E.g., if all the other 
parameters are nonzero then we have a 
D6--$(\rm D4)^{3}$--$(\rm D2)^{3}$--D0--F1--W
 bound state\footnote{Here $(\rm D4)^{3}$ means that there are three 
D4--branes in the bound state.}. 
In this case the metric can be 
rewritten in the following way (with $\theta^{01}=-\theta^{12}=\theta$ and 
$h=1$):
\begin{eqnarray}\label{W}
d\tilde{s}^{2}&=&H^{-1/2}\Big[(-dx^{0}+dx^{2})(dx^{0}+dx^{2})+(dx^{1})^{2}-
H^{-1}\theta^{2}(-dx^{0}+dx^{2})^{2}\nonumber\\
&+&\frac{1}{h_{34}}((dx^{3})^{2}+(dx^{4})^{2})+\frac{1}{h_{56}}
((dx^{5})^{2}+(dx^{6})^{2})+H(dr^{2}+r^{2}d\Omega_{2}^{2})\Big],\ 
 \end{eqnarray}
which implies that there is a wave in the 2nd direction. This can be seen by
going to light-like coordinates $x^{\pm}=\frac{1}{\sqrt{2}}(x^{2}\pm x^{0})$.

Instead if  
$|\theta^{01}|>|\theta^{12}|$ one can perform a Lorentz transformation to a 
frame where $\theta^{01}\neq 0$ and $\theta^{12}=0$\footnote{We thank P. 
Sundell for pointing this out to us. See also \cite{wu,LLNCYM}.}. Then if all
 the other 
parameters are nonzero, we have a D6--$(\rm D4)^{2}$--D2--F1 bound state. 
Finally, 
if $|\theta^{01}|<|\theta^{12}|$, one can perform a Lorentz transformation 
to a frame where $\theta^{01}=0$ while $\theta^{12}\neq 0$, which gives a 
 D6--$(\rm D4)^{3}$--$(\rm D2)^{3}$--D0 bound state provided also 
$\theta^{34}$ and $\theta^{56}$ are nonzero. In Table 1 we 
show in 
which directions the different branes are oriented (for the 
D6--$(\rm D4)^{3}$--$(\rm D2)^{3}$--D0 bound state).

\begin{table}
	\begin{center}
	\begin{tabular}{|c|c|c|c|c|c|c|} \hline
	Brane & $x^{1}$ & $x^{2}$ & $x^{3}$ & $x^{4}$ & $x^{5}$ & $x^{6}$ \\ 
\hline
	D6 & X & X & X & X & X & X \\ \hline
	D4 & X & X & X & X & $-$ & $-$ \\ \hline
	D4 & $-$ & $-$ & X & X & X & X \\ \hline
	D4 & X & X & $-$ & $-$ & X & X \\ \hline
	D2 & X & X & $-$ & $-$ & $-$ & $-$ \\ \hline
	D2 & $-$ & $-$ & X & X & $-$ & $-$ \\ \hline
	D2 & $-$ & $-$ & $-$ & $-$ & X & X \\ \hline
	D0 & $-$ & $-$ & $-$ & $-$ & $-$ & $-$ \\ \hline
	
	\end{tabular}
	\end{center}
	\caption{The D6--$(\rm D4)^{3}$--$(\rm D2)^{3}$--D0 bound state}
	\label{D64442220}
\end{table}

Using T-duality, all these bound states can be used to obtain bound 
states in type IIB with a D5--brane (see \cite{cederwall1} for the complete 
$(p,q)$ 5--brane bound states) or a D7--brane as the highest rank brane.
However, T-duality of the deformed D6--brane solution (\ref{D6W}) can not be 
used in order to obtain a D7--brane bound state, with a rank 8 
($\theta^{i,i+1}$ $i=0,2,4,6$) deformation. This 
D7--$(\rm D5)^{3}$--$(\rm D3)^{3}$--D1--F1 
bound state is therefore included in Appendix A. 

\subsection{Non--commutative theories}

In the three different cases above it is possible to take near horizon limits, 
e.g., if one wants to obtain supergravity duals of non--commutative open brane
 (or gauge) theories. In the first case, with 
$\theta^{01}=-\theta^{12}=\theta$ (which correspond to a light-like 
deformation, since $B_{-1}\neq 0$ and $B_{+1}=0$)
and $\theta^{i,i+1}\neq 0$ ($i=3,5$), one has to take a magnetic near 
horizon 
limit \cite{per}, i.e., keeping the following quantities fixed:\footnote{
This limit gives a finite supergravity lagrangian, since 
$d\tilde{s}^{2}/\alpha'$, 
$\tilde{B}_{2}/\alpha'$ and $\tilde{C}_{q}/(\alpha')^{q/2}$ are kept fixed in 
the $\alpha'\rightarrow 0$ limit.}
\begin{equation}\label{mag}
x^{\mu}\ ,\quad u=\frac{r}{\alpha'}\ ,\quad \tilde{\theta}=\alpha'\theta\ ,
\quad
\tilde{\theta}^{i,i+1}=\alpha'\theta^{i,i+1}\ ,\quad 
g^{2}_{\rm YM}=g(\alpha')^{3/2}\ ,
\end{equation}
in the $\alpha' \rightarrow 0$ limit.
Taking this limit for the $\theta^{01}=-\theta^{12}$ case, gives the 
supergravity dual of NCYM theory with 
light-like ($\Theta^{+1}=\sqrt{2}\tilde{\theta}$, $\Theta^{-1}=0$), as well as 
space-space non--commutativity ($\Theta^{i,i+1}=\tilde{\theta}^{i,i+1}$, 
$i=3,5$)\footnote{For other NCYM theories with light-like non--commutativity
 see e.g. \cite{al1,LLNCYM}.}. This can be seen from the metric (\ref{W}) by
 going to light-like coordinates $x^{\pm}=\frac{1}{\sqrt{2}}(x^{2}\pm x^{0})$.

The above limit can also be used in all other cases when 
$\theta^{01}=0$. Then one obtains supergravity duals of NCYM with 
space-space non--commutativity ($\Theta^{i,i+1}=\tilde{\theta}^{i,i+1}$, 
$i=1,3,5$), which was shown in \cite{al}. 

In the case when 
$\theta^{01}=\theta^{12}=\theta^{34}=0$ and \mbox{$\theta^{56}\neq 0$}, 
one can also 
interpret the supergravity solution as being dual to an open D4--brane theory 
(since there is a critical RR 5-from in the solution, which is responsible for
the finite tension $T=\frac{1}{\ell^{5}_{D4}}=\frac{1}{g^{2}_{\rm YM}
\tilde{\theta}^{56}}$ of an open D4--brane),   
containing light open D4--branes analogous to the open D$q$--brane theories
in \cite{ODq,solo1}\footnote{In \cite{solo1} it was shown that the 
supergravity dual of $(q+3)$--dimensional NCYM (with one non-zero $\theta$) is
 identical to the supergravity dual of an open D$q$--brane theory, provided
 one makes the identification (25) in \cite{solo1}. Note that the case $q=4$
was not discussed in \cite{solo1}, but that equation (25) is still valid in 
this case. For a more thorough investigation of open 
D$q$--brane theories ($q$=0,1,2,3), see \cite{solo1}.}. Using the results in
 Section 4 in \cite{solo1} 
 (defining the coupling in a similar fashion) gives that the open D4--brane 
theory, with length-scale $\ell_{\rm D4}$ and coupling constant 
$G^{2}_{\rm D4}$, 
and the 7--dimensional NCYM theory (with only $\tilde{\theta}^{56}\neq 0$) are 
related as follows:
\begin{equation}\label{openD4}
g^{2}_{\rm YM}=G_{\rm D4}^{4/5}\ell_{\rm D4}^{3}\ ,\quad 
\Theta^{56}=\tilde{\theta}^{56}=G_{\rm D4}^{-4/5}\ell_{\rm D4}^{2}\ .
\end{equation} 
This implies that 7-dimensional NCYM (with only $\tilde{\theta^{56}}\neq 0$) 
arises in the limit of small $G^{2}_{D4}$ and small length scale 
$\ell_{D4}$ (keeping $\tilde{\theta}^{56}=G_{\rm D4}^{-4/5}\ell_{\rm D4}^{2}$ 
fixed), which is the weakly coupled low energy limit of the open 
D4--brane theory, where we expect to have an effective field theory 
description.

When $\theta^{01}\neq 0$ and $\theta^{12}=0$ 
 one has to take an
 electric near horizon limit, i.e., keeping the following quantities 
fixed \cite{solo,solo1}\footnote{To be more 
precise, $\theta^{01}=k$ where $k$ is the constant in the harmonic function 
$H$. In this note we always set $k=1$.}:
\begin{equation}\label{el}
\tilde{x}^{\mu}=\sqrt{\frac{\alpha'_{\rm eff}}{\alpha'}}x^{\mu}\ ,\quad 
\tilde{r}=\sqrt{\frac{\alpha'_{\rm eff}}{\alpha'}}r\ ,\quad \theta^{01}=1\ ,
\quad \theta^{i,i+1}\ ,\quad g\ ,
\end{equation}
in the $\alpha'\rightarrow 0$ limit.
Note that this limit is very different from (\ref{mag}) since the limit is 
obtained by making the electric field critical (i.e., $\theta^{01}=1$). In the
 $\theta^{01}\neq 0$ and $\theta^{12}=0$ case this   
limit gives the supergravity dual of 7--dimensional NCOS with the following 
open string quantities:
\begin{eqnarray}
\frac{G_{\mu\nu}}{\alpha'}&=&\frac{1}{\alpha'_{\rm eff}}H^{-1/2}\eta_{\mu\nu}
\ ,\quad 
G^{2}_{OS}=gH^{-3/4}\ ,\nonumber\\
\Theta^{01}&=&\alpha'_{\rm eff} ,\quad 
\Theta^{34}=\theta^{34}\alpha'_{\rm eff}\ ,
\quad \Theta^{56}=\theta^{56}\alpha'_{\rm eff}\ ,
\end{eqnarray}
which means that we have both space-time and space-space non--commutativity.

However, since all the above theories are world volume theories of 
a deformed D6--brane, it is possible that some or all of them do not 
decouple from the bulk gravity. This is a possibility since the undeformed 
super-Yang-Mills theory on the D6--brane does not decouple \cite{mal}. 
In e.g. 
\cite{LLNCYM} it was suggested that NCYM theory with light-like 
non--commutativity does not decouple, and in \cite{oz} it was shown that 
NCYM with space-space non-commutativity is not a decoupled theory.
For NCOS on the other hand it was argued in \cite{sei,witten} that NCOS 
 decouple for all D$p$--branes with $p<7$. 
Whether the open D4--brane theory decouples or not is unclear, but since it has the 
same supergravity dual as NCYM (with only $\theta^{56}\neq 0$) it indicates 
 that it is not a decoupled theory. A further investigation of this will be 
presented elsewhere.

\section{M-theory bound states}

Next, we lift the type IIA bound state solutions (\ref{D6W})  
to obtain half-supersymmetric M-theory bound state solutions\footnote{Note 
that uplift to 11 dimensions does not reduce supersymmetry.}.
Lifting the type IIA solution (\ref{D6W}) to 11 dimensions on a circle with 
radius $R_{11}=g\sqrt{\alpha'}$ gives\footnote{
We use the conventions used in \cite{solo1}. Note also that we do not include 
the auxiliary 6-form.}

\begin{eqnarray} \label{KK6}
ds^{2}_{11}&=&(hh_{34}h_{56})^{1/3}\Big[\frac{1}{h}(-h_{12}(dx^{0})^{2}+
(dx^{1})^{2}+h_{01}(dx^{2})^{2})-2\frac{1}{Hh}\theta^{01}\theta^{12}
dx^{0}dx^{2}\nonumber\\
& &+\frac{1}{h_{34}}((dx^{3})^{2}+(dx^{4})^{2})+\frac{1}{h_{56}}
((dx^{5})^{2}+(dx^{6})^{2})+H(dr^{2}+r^{2}d\Omega_{2}^{2})\nonumber\\
& &+\frac{1}{Hhh_{34}h_{56}}\Big(dx^{10}-R\epsilon_{1}+
\frac{\theta^{12}\theta^{34}\theta^{56}}{H}
dx^{0}+\frac{\theta^{01}\theta^{34}\theta^{56}}{H}dx^{2}\Big)^{2}\Big]\ ,    
 \nonumber \\
A_{3}&=&\frac{\theta^{34}\theta^{56}}{Hh} dx^{0}
\wedge dx^{1}\wedge dx^{2}+\frac{\theta^{12}\theta^{56}}{Hh_{34}} dx^{0}
\wedge dx^{3}\wedge dx^{4}+\frac{\theta^{12}\theta^{34}}{Hh_{56}} dx^{0}
\wedge dx^{5}\wedge dx^{6} \nonumber\\
& &+\frac{\theta^{01}\theta^{34}}{Hh_{56}} dx^{2}\wedge dx^{5}\wedge 
dx^{6}+\frac{\theta^{01}\theta^{56}}
{Hh_{34}} dx^{2}\wedge dx^{3}\wedge dx^{4} 
\\
& &+\Big({\theta^{01}\over Hh}dx^0\wedge dx^1
-{\theta^{12}\over  Hh}dx^1\wedge dx^2
-{\theta^{34}\over  Hh_{34}}dx^3\wedge dx^4 \nonumber\\
& &-{\theta^{56}\over  Hh_{56}}dx^5\wedge dx^6\Big)\wedge (dx^{10}-
R\epsilon_{1})
\ , \quad H=1+\frac{R}{r}\ .\nonumber
\end{eqnarray}
Depending on the values of the $\theta$ parameters, we obtain several different
 bound states, e.g., $\theta^{01}=0$, and $\theta^{i,i+1}\neq 0$ ($i=1,3,5$)
 gives a half-supersymmetric KK6--$(\rm M5)^{3}$--$(\rm M2)^{3}$--MW bound 
state (i.e., uplifting of the 
D6--$(\rm D4)^{3}$--$(\rm D2)^{3}$--D0 bound state). 

With $\theta^{01}=\pm \theta^{12}$ and 
$\theta^{i,i+1}\neq 0$ ($i=1,3,5$) we instead obtain a half-supersymmetric
KK6--$(\rm M5)^{3}$--$(\rm M2)^{4}$--$(\rm MW)^{2}$ bound state 
(which is uplifting to M-theory of the 
D6--$(\rm D4)^{3}$--$(\rm D2)^{3}$--D0--F1--W bound state). In this case the
 metric can be 
rewritten in the following way (with $\theta^{01}=-\theta^{12}=\theta$ and 
$h=1$):
\begin{eqnarray}\label{W1}
d\tilde{s}^{2}_{11}&=&(h_{34}h_{56})^{1/3}\Big[(-dx^{0}+dx^{2})(dx^{0}+dx^{2})
+(dx^{1})^{2}-H^{-1}\theta^{2}(-dx^{0}+dx^{2})^{2}\nonumber\\
& &+\frac{1}{h_{34}}((dx^{3})^{2}+(dx^{4})^{2})+\frac{1}{h_{56}}
((dx^{5})^{2}+(dx^{6})^{2})+H(dr^{2}+r^{2}d\Omega_{2}^{2})\nonumber\\
& &+\frac{1}{Hh_{34}h_{56}}\Big(dx^{10}-R\epsilon_{1}+
\frac{\theta\theta^{34}\theta^{56}}{H}(-dx^{0}+dx^{2})\Big)^{2}\Big]\ ,
\end{eqnarray}
where the waves are in the 2nd and 10th directions.

Finally, if we have $\theta^{01}\neq 0$, $\theta^{12}=0$ and 
$\theta^{i,i+1}\neq 0$,
 $i=3,5$, we obtain a KK6--$(\rm M5)^{2}$--$(\rm M2)^{2}$ bound state 
(i.e., uplifting of D6--$(\rm D4)^{2}$--D2--F1).

\section{Discussion}

In this note we have obtained type IIA bound states, containing D6, D4, D2,
 D0, F1--branes and waves, using $O(7,7)$ transformations on D6--branes. 
Lifting 
these bound states to M-theory gives bound states containing KK6, M5, 
M2--branes and M--waves. In an Appendix we have also included a type IIB 
D7--$(\rm D5)^{3}$--$(\rm D3)^{3}$--D1--F1 bound state.
For the type IIA bound states we take electric 
(\ref{el}) and magnetic (\ref{mag}) near horizon limits, which give us 
supergravity duals of NCOS (with $\Theta^{01}=\alpha'_{\rm eff}$,
 $\Theta^{34}=\theta^{34}\alpha'_{\rm eff}$ and 
$\Theta^{56}=\theta^{56}\alpha'_{\rm eff}$) and NCYM (with light-like and 
space-space non-commutativity). In the case of NCYM with only 
$\tilde{\theta}^{56}\neq 0$ 
we argue that the supergravity dual can also be interpreted as supergravity 
dual of a theory containing light open D4--branes, where the relation between 
the open D4--brane theory and NCYM is given in (\ref{openD4}). This suggest 
that NCYM (with only $\tilde{\theta}^{56}\neq 0$) is the low energy limit of
 the 
open D4--brane theory, i.e., the open D4--brane theory is the UV completion 
of 7-dimensional NCYM. 

In \cite{oz} it was shown that NCYM on the D6--brane does not decouple. For  
NCOS on the other hand it was argued in \cite{sei,witten} that NCOS decouple
 for all D$p$--branes with
$p<7$. The possible decoupling of the open D4--brane theory has to be 
investigated further. However, the fact that it has the same supergravity dual
as NCYM (with only $\tilde{\theta}^{56}\neq 0$) might indicate that it is
 not a decoupled theory.  

Next, to study the strong coupling of NCOS (with only $\Theta^{01}\neq 0$), 
one has to lift the D6-F1 bound state to M-theory, which gives the KK6-M2 
bound state, and take the appropriate near horizon limit\footnote{The easiest 
way 
to obtain this supergravity dual is to lift the D6-F1 bound state after one 
has taken an electric near horizon limit.}. What complicates this set up is 
that the KK6--brane is in the $x^{1}-x^{6}$ directions while the M2--brane 
has one direction in the periodic $x^{10}$ direction, i.e., transverse to the 
KK6--brane. For large $\tilde{r}$ the metric approaches a smeared membrane, 
which in the case of the near horizon region of the M5-M2 bound state 
indicates the existence of an open membrane theory. In this case, however, 
one can not adopt the same interpretation since one of the membrane directions 
is transverse to the KK6--brane. It remains to be seen what the strong 
coupling limit of 7-dimensional NCOS is, but we do not expect it to be an 
open membrane theory. 

As the next step, it would be interesting to find out if NCOS and/or the open 
D4--brane theories really are decoupled or not, and also what their strong 
coupling limits are. One can also further investigate 
the M-theory bound states KK6--$(\rm M5)^{3}$--$(\rm M2)^{3}$--MW and 
KK6--$(\rm M5)^{3}$--$(\rm M2)^{4}$--$(\rm MW)^{2}$, in order to find dual 
theories. However, these theories are probably not decoupled, since they
 are strong
 coupling limits of NCYM with space-space non--commutativity and NCYM with 
space-space as well as light-like non--commutativity, respectively.
 
Deformations of a type IIA D8--brane should also be possible, but since the 
D8--brane only exists in massive type IIA, one would have to adjust the 
deformation procedure in order to include these deformations. This has not yet 
been done, but we expect that it is possible to deform the D8--brane with 
e.g. a magnetic rank 8 deformation $\theta^{i,i+1}$, $i=1,3,5,7$, which 
should give a half-supersymmetric 
D8--$(\rm D6)^{4}$--$(\rm D4)^{6}$--$(\rm D2)^{4}$--D2 bound state.

\vspace{0.5cm}
\Large {\bf Acknowledgments}
\normalsize
\vspace{0.3cm}\\
We thank M. Cederwall, U. Gran, M. Nielsen, 
B.E.W. Nilsson and P. Sundell for  discussions and comments.

\appendix

\section{Deformation of the D7--brane}

In this appendix we will deform the D7--brane with a rank 8 deformation, by 
turning on $\theta^{01},\theta^{23},\theta^{45}$ and $\theta^{67}$. Note that
 a rank 8 deformation is not possible for the D6-brane. The resulting 
D7--brane bound state can therefore not be obtained from 
the deformed D6--brane (\ref{D6W}), using T-duality. The undeformed D7--brane 
configuration is given by (see e.g. \cite{ort})\footnote{Note that 
we do not include the auxiliary RR $C_{0}$ potential. For an introduction to 
\mbox{7--branes} and their transformations properties under $SL(2,Z)$, see 
\cite{ort}.}
  
\begin{eqnarray}\label{7}
ds^2&=&H^{-1/2}(-(dx_{0})^{2}+\ldots +(dx_{7})^{2}) + H^{1/2}(dr^{2}+r^{2}
d\Omega^{2}_{1})\ ,\nonumber\\
C_{8}&=&\frac{1}{gH}dx^0\wedge\cdots dx^7\ ,\\ 
e^{2\phi}&=&g^{2}H^{-2}\ ,\quad H=k+\frac{g}{2\pi}\log{\frac{r}{R}}\ ,
\nonumber \end{eqnarray}
where $k$ is an arbitrary constant.
Deforming the D7--brane 
 with $\theta^{01},\theta^{23},\theta^{45}$ and $\theta^{67}$ gives
\begin{eqnarray} \label{D7M}
ds^{2}&=&H^{-1/2}\Big(\frac{1}{h_{01}}(-dx^{2}_{0}+dx^{2}_{1})
+\frac{1}{h_{23}}(dx^{2}_{2}+dx^{2}_{3})+\frac{1}{h_{45}}(dx^{2}_{4}
+dx^{2}_{5})\nonumber\\
& &+\frac{1}{h_{67}}(dx^{2}_{6}+dx^{2}_{7})\Big)+H^{1/2}(dr^{2}+
r^{2}d\Omega^{2}_{1})\ , \nonumber\\
e^{2\tilde{\phi}}&=&\frac{g^{2}H^{-2}}{h_{01}h_{23}h_{45}h_{67}} \ ,\nonumber\\
\tilde{B}&=&{\theta^{01}\over Hh_{01}}dx^0\wedge dx^1
-{\theta^{23}\over Hh_{23}}dx^2\wedge dx^3
-{\theta^{45}\over Hh_{45}}dx^4\wedge dx^5
-{\theta^{67}\over Hh_{67}}dx^6\wedge dx^7\ ,\nonumber\\
g\tilde{C}_8&=&{1 \over Hh_{01}h_{23}h_{45}h_{67}}dx^0\wedge
\cdots \wedge dx^7\ ,\\
g\tilde{C}_6&=& -\frac{\theta^{01}}{Hh_{23}h_{45}h_{67}}
dx^{2}\wedge\cdots \wedge dx^{7}-\frac{\theta^{23}}
{Hh_{01}h_{45}h_{67}}dx^{0}\wedge dx^{1}\wedge dx^{4}\wedge\cdots
\wedge dx^{7}\nonumber\\
& &-\frac{\theta^{45}}{Hh_{01}h_{23}h_{67}}
dx^{0}\wedge\cdots\wedge dx^{3}\wedge dx^{6}\wedge dx^{7}
-\frac{\theta^{67}}
{Hh_{01}h_{23}h_{45}}dx^{0}\wedge\cdots \wedge dx^{5}\ ,\nonumber\\
g\tilde{C}_{4}&=&\frac{\theta^{01}\theta^{23}}{Hh_{45}h_{67}}
 dx^{4}\wedge dx^{5}\wedge dx^{6}\wedge dx^{7}+\frac{\theta^{01}
\theta^{45}}{Hh_{23}h_{67}} dx^{2}
\wedge dx^{3}\wedge dx^{6}\wedge dx^{7}\nonumber\\
& &+\frac{\theta^{01}\theta^{67}}
{Hh_{23}h_{45}} dx^{2}\wedge dx^{3}\wedge dx^{4}\wedge dx^{5}
+\frac{\theta^{23}\theta^{45}}{H
h_{01}h_{67}} dx^{0}\wedge dx^{1}\wedge dx^{6}\wedge dx^{7}\nonumber\\
& &+\frac{\theta^{23}
\theta^{67}}{Hh_{01}h_{45}} dx^{0}\wedge dx^{1}\wedge 
dx^{4}\wedge dx^{5}+\frac{\theta^{45}
\theta^{67}}{Hh_{01}h_{23}} dx^{0}\wedge dx^{1}\wedge 
dx^{2}\wedge dx^{3}\ ,\nonumber\\
g\tilde{C}_{2}&=&-\frac{\theta^{01}\theta^{23}\theta^{45}}
{Hh_{67}}dx^{6}\wedge dx^{7}-\frac{\theta^{01}
\theta^{23}\theta^{67}}
{Hh_{45}}dx^{4}\wedge dx^{5}\nonumber\\
& &-\frac{\theta^{01}\theta^{45}\theta^{67}}
{Hh_{23}}dx^{2}\wedge dx^{3}-\frac{\theta^{23}\theta^{45}\theta^{67}}
{Hh_{01}}dx^{0}\wedge dx^{1}\ ,\nonumber \\
g\tilde{C}_{0}&=&\frac{\theta^{01}\theta^{23}\theta^{45}\theta^{67}}
{H}\ ,\nonumber\\
\mbox{with}& & h_{i,i+1}=1+(\theta^{i,i+1})^{2}H^{-1}\ , \quad i=2,4,6\ ,
\quad h_{01}=1-(\theta^{01})^{2}H^{-1}\ . \nonumber 
\end{eqnarray}
 
For $\theta^{i,i+1}\neq 0$ ($i=0,2,4,6$) we obtain a 
D7--$(\rm D5)^{3}$--$(\rm D3)^{3}$--D1--F1 bound state. If $\theta^{01}=0$ 
the F1--string will no longer be included. Instead, if one magnetic $\theta$ 
vanishes 
the D1--brane is removed, while two magnetic $\theta'$s equal to zero removes
 the D1 and D3--branes and finally, all three equal to zero removes the D1, D3
 and D5--branes.

\end{document}